\documentclass[twocolumn, pre]{revtex4-1}
\usepackage{graphicx} 
\usepackage{amsmath}
\usepackage{bm}
\usepackage{amssymb}
\usepackage{natbib} 

\begin{document}
\title{Anomalous dynamics of an elastic membrane in an active fluid}

\author{S. A. Mallory$^1$, C. Valeriani$^{2}$, A. Cacciuto$^1$}
\email{ac2822@columbia.edu}
\affiliation{$1$Department of Chemistry, Columbia University\\ 3000 Broadway, New York, NY 10027\\ }
\affiliation{$^{2}$Departamento de Fisica Aplicada I, Facultad de Ciencias Fisica, Universidad Complutense de Madrid, 28040 Madrid, Spain}

\begin{abstract}
Using numerical simulations, we characterized the behavior of an elastic membrane immersed in an active fluid.  Our findings reveal a nontrivial folding and re-expansion of the membrane that is controlled by the interplay of its resistance to bending and the self-propulsion strength of the active components in solution. We show how flexible membranes tend to collapse into multi-folded states, whereas stiff membranes oscillates between an extended configuration and a singly folded state.  This study provides a simple example of how to exploit the random motion of active particles to perform mechanical work at the micro-scale.
\end{abstract}

\maketitle

\noindent \textit{Introduction} -- Suspensions of bacteria and synthetic active particles offer a novel approach to manipulating matter at the microscale. Not only do passive micro-components in an active fluid display unusual transport properties \cite{Wang17744, Leptos198103, Dunkel4268,Morozov2748, Mallory032309, molina1389,douglass834, mino1469, kasyap081901,kaiser158101, wu3017,mino048102,eckhardt96,jepson041002,shklarsh786,angelani113017,takagi1784}, but active fluids can mediate
a new set of  effective interactions between passive elements \cite{kaiser1411.0964, harder194901, ni018302, ray013019, angelani138302,das198301,krafnick022308}, providing an extra handle in material design. In addition, active fluids have been used to power primitive micromachines \cite{Angelani048104,lambert168102,Sokolov969, DiLeonardo9541,koumakis,schwarz4052}.  The origin of this phenomenology is derived from the unique pressure (or stress) gradients generated by active fluids \cite{grosberg1502.08034,solon1412.3952,solon1412.5475,mallory052303,fily5609,yang6477,takatori1411.5776,ginot011004,takatori028103,takatori9433}.

Although, much of the effort in this field has focused on the interaction of active fluids with rigid, passive microcomponents, there has been some work on the behavior of flexible objects in an active fluid. Both flexible\cite{harder062312,kaiser044903} and semi-flexible polymers\cite{harder062312} confined to a two-dimensional active bath have been shown to exhibit dynamic and scaling behavior that is much richer than that expected for polymers in a thermal bath.  Fully flexible polymers\cite{kaiser044903} display a non-universal Flory scaling behavior as well as an anomalous chain swelling, while semi-flexible polymers\cite{harder062312} display a dynamic collapse and re-expansion when immersed in an active bath at different values of the propelling forces.  These results suggest that the driven nature of an active fluid can be used to control the  shape of flexible microcomponents. This can be thought of as a microscopic joint or clamp where the mechanical action (folding) of stiff fibers can be induced by increasing the self-propulsion of the active components in solution.

Much of the phenomenological behavior observed for polymers is, however, specific to two-dimensional (or quasi-two dimensional) systems, where active particles can be confined with relative ease within the bends of a polymer, and generate significant pressure gradients capable of driving its collapse.  The scaling behavior and statistical properties of a polymer, when embedded in a three-dimensional active fluid, do not differ significantly from those of a polymer in a thermal bath. This would not be the case when replacing the polymer with a flexible two dimensonal surface, which is the subject of this work.

Here, we study the behavior of an extended elastic membrane, a  natural generalization of a linear polymer chain to an intrinsically two-dimensional structure, suspended in a three dimensional active bath. Physical examples of such materials include graphite-oxide and graphene sheets\cite{meyer60,stankovich282,wen426,spector2867}, cross-polymerized biological membranes\cite{fendler3}, and the cytoskeleton of red blood cells\cite{elgsaeter1217,schmidt952}.  Polymerized membranes have been studied intensively in the last few decades (see \cite{bowick255} and references therein for a review on the subject) and display a phenomenological behavior that is far richer in complexity than that observed in linear polymers. 
Using numerical simulations, we explore the mechanical properties and conformational behavior of an elastic membrane immersed in an active fluid for different strengths of the bath activity and for different values of its bending rigidity.  Our findings reveal a nontrivial folding and re-expansion of the membrane that is controlled by the interplay of its resistance to bending and the self-propulsion strength of the active components in solution. We show how flexible membranes tend to collapse into multi-folded states, whereas stiff membranes oscillates between an extended configuration and a singly folded state.\\

\begin{figure}[h!]
\includegraphics[width=0.5\textwidth]{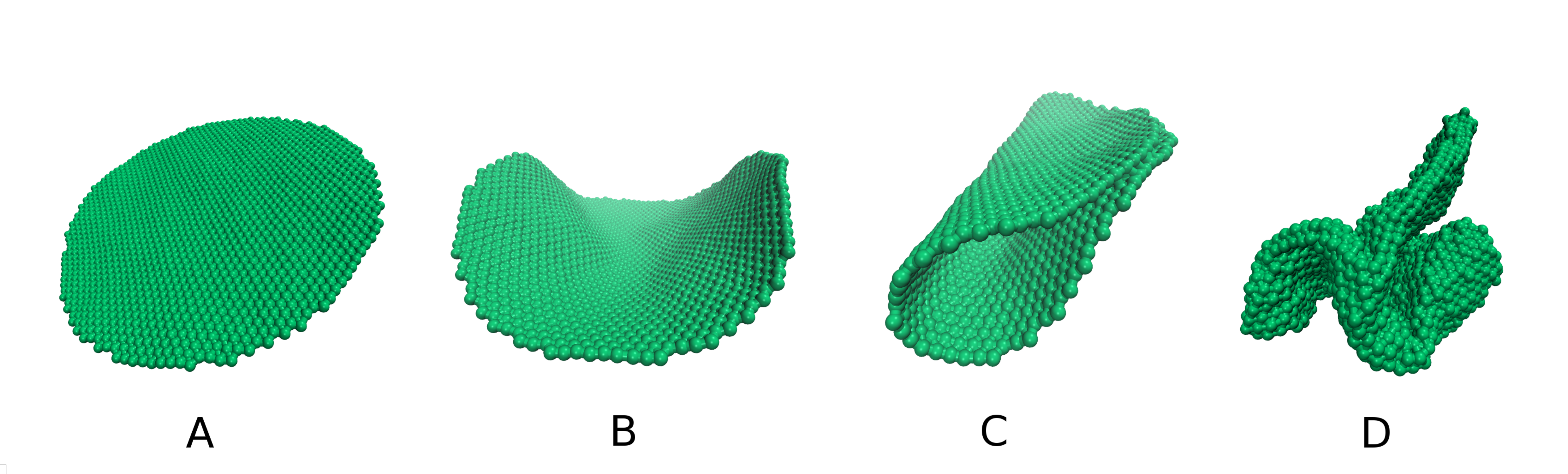}
\caption{\label{schematic} {(Color online)  Reference snapshots showing the various conformations taken by the membrane as the bath activity and bending rigidity is varied. (A) flat/extended, (B) bent, (C) single folded, and (D) multi-folded.}}
\end{figure}

\pagebreak

\noindent\textit {Methods} -- The elastic membrane is modeled using a standard triangulated mesh with hexagonal symmetry \cite{kantor2774}. The mesh is composed of $N_m$ nodes which are arranged in a circular geometry with diameter $d$. Self-avoidance of the surface is imposed by placing a purely repulsive particle of diameter $\sigma$ and mass $m$ at each node of the mesh. Every particle on the surface is bonded to its first neighbor via a harmonic potential

\begin{equation}
U_{stretching}=\kappa_s(r-r_0)^2
\end{equation}

\noindent where $\kappa_s$ is the spring constant, $r$ the distance between two neighboring particles, and $r_0 = 2^{1/6} \sigma$ the equilibrium bond length between two particles. The bending rigidity of the membrane is implemented using a dihedral potential between adjacent triangles of the mesh:

\begin{equation}
U_{bending}=\kappa_b(1+\cos \phi)
\end{equation}

\noindent where $\phi$ is the dihedral angle between opposite vertices of any two triangles sharing an edge and $\kappa_b$ is the bending constant.

Each active bath particle is a sphere with mass $m$, diameter $\sigma$, and undergoes overdamped Langevin dynamics at a constant temperature $T$. Self-propulsion is introduced through a directional force of constant magnitude $|F_a|$ and is directed along a predefined orientation vector $\boldsymbol{n}$ which passes through the origin of each particle and connects its poles. The equations of motion of an individual particle are given by the coupled Langevin equations

\begin{align}
m \ddot{ \pmb{r}} & =-\gamma \dot{ \pmb{r}}-\partial_{\pmb{r}} V+|F_a|\pmb{n}+ \sqrt{2 \gamma^2 D} \pmb{\xi}(t) \\
\dot{\pmb{n}} & = \sqrt{2D_r} \pmb{\xi_R(t)} \times \pmb{n}
\end{align}

\noindent where $\gamma$ is the drag coefficient, $V$ the interparticle potential, and $D$ and $D_r$ are the translational and rotational diffusion constants respectively, satisfying the relation $D_r=(3D)/\sigma^2$. The translational diffusion constant $D$ is related to the temperature $T$ via the Stokes-Einstein relation $D=k_BT/\gamma$. The typical solvent induced Gaussian white noise terms for both the translational and rotational motion are characterized by $\langle \xi_i(t) \rangle = 0$ and $\langle \xi_i(t) \cdot \xi_j(t') \rangle = \delta_{ij}\delta(t-t')$ and $\langle \xi_{Ri}(t)\rangle = 0$ and $\langle \xi_{Ri}(t) \cdot \xi_{Rj}(t') \rangle = \delta_{ij}\delta(t-t')$, respectively. Each node of the membrane also undergoes over-damped Langevin dynamics at a constant temperature $T$ where the equations of motion are given by Eqs. (3) and (4) while letting $|F_a|=0$. The interactions between any two particle in the systems (membrane nodes or active components) are purely repulsive and are given by the Weeks-Chandler-Andersen (WCA) potential
\begin{equation}
U(r_{ij})=4 \epsilon \left[ \left( \frac{\sigma}{r_{ij}} \right)^{12}- \left( \frac{\sigma}{r_{ij}} \right)^{6}+\frac{1}{4} \right] 
\label{eq:LJ_V}
\end{equation}  
\noindent with a range of action extending up to $r_{ij}=2^{1/6}\sigma$. Here $r_{ij}$ is the center-to-center distance between any two particles $i$ and $j$, and $\epsilon$ is the interaction energy. 

\begin{figure}[b!]
\includegraphics{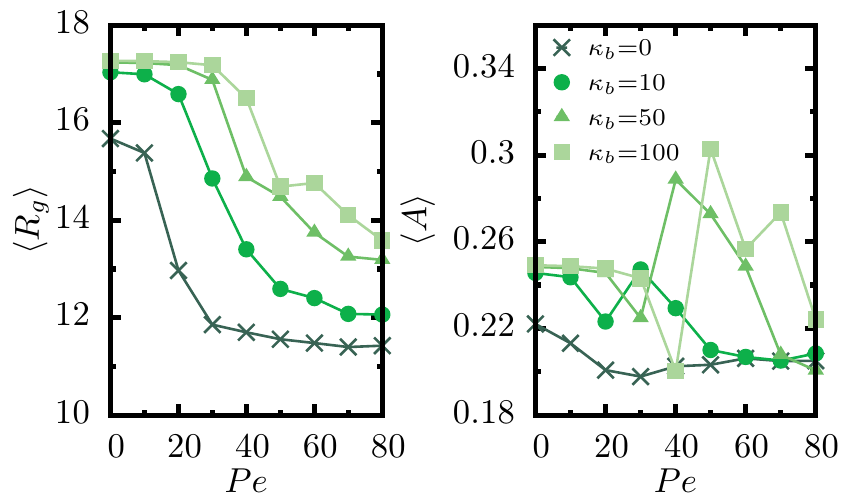}
\caption{\label{avg_Rg} {(Color online) The time average of the radius of gyration $\langle R_g \rangle$ (left)  and asphercity $\langle A \rangle$ (right) as a function of $Pe$ for membranes of different bending rigidities $\kappa_b$.}}
\end{figure}

\begin{figure*}[t!]
\includegraphics{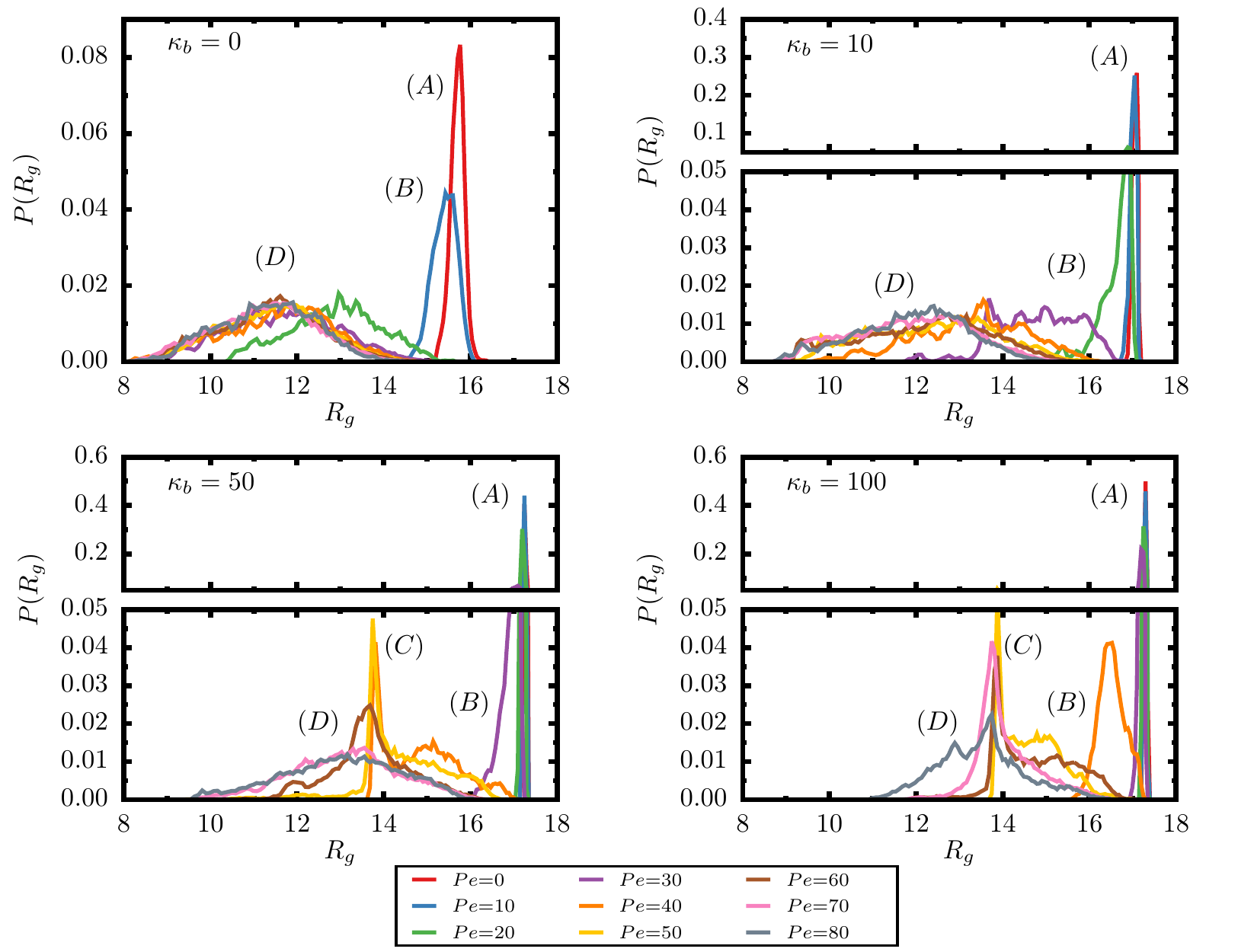}
\caption{\label{Rg} {(Color online) Probability distribution of the radius of gyration $P(R_g)$ for membranes of increasingly large Peclet numbers.  
Different panels show the results for different bending rigidities of the membrane as indicated. The letters in parenthesis indicate the 
corresponding configurations as shown in Fig.~\ref{schematic}.
}}
\end{figure*}

Using the numerical package LAMMPS \cite{plimpton1}, all simulations were carried out in a periodic box of dimension $L=100$ with $T= \epsilon=m=\sigma=\tau=1$ and $\gamma=10\tau^{-1}$(here $\tau$ is the dimensionless time). The drag coefficient $\gamma$ was chosen to be sufficiently large such that the motion of the particles is effectively overdamped. The number of active particles considered in our simulations was $N=10,000$ giving an active particle density of $\rho=N/V=0.01$. While, the number of nodes in the membrane was set to $N_m=1700$ resulting in a membrane with diameter $d \approx 46 \sigma$. For each simulation, the membrane was initialized in a flat configuration and the simulation was run for a minimum of $1 \times 10^8 \tau$ time steps. All quantities in this investigation are given in reduced Lennard-Jones units and for convenience, we refer to the activity of the bath in terms of the dimensionless Peclet number $Pe=|F_a| \sigma/(\gamma D)=|F_a| \sigma/(k_BT)$. \\

\noindent \textit{Results} -- To characterize and quantitatively differentiate between the different conformations of the membrane, we employ 
two well established shape parameters: the membrane asphericity $A$ and its radius of gyration $R_g$. 
Following Rudnick et al. \cite{rudnickL191}, we define asphericity as the rotational invariant
\begin{equation}
A=\frac{\sum_{i>j} (\lambda_i-\lambda_j)^2 }{(\sum_i\lambda_i)^2 }
\end{equation}
where $\lambda_i,\lambda_j$ are the $i$th and $j$th eigenvalues of the inertia tensor, and $i,j \in (1,2,3)$. A fully symmetric object such as a sphere (i.e. $\lambda_1 = \lambda_2 = \lambda_3$), will have a value of asphericity equal to zero. At the opposite extreme, a thin rod (i.e. $\lambda_1>0,\lambda_2 =\lambda_3=0$) will have an asphericity equal to one. We are predominately concerned with a planar circular membrane whose conformation oscillates between an extended state ($A \approx 0.25$), a singly folded state ($A \approx 0.5$), and an ensemble of multi-folded crumpled states (that are the most symmetric)  which are characterized by the smallest values of $A$.  Representative snapshots of  these configurations are shown in Fig.~\ref{schematic}.  The second configurational parameter, the radius of gyration of the membrane, 
is simply the trace of the inertia tensor given by
\begin{equation}
R^2_g=\sum_i\lambda_i 
\end{equation} 

\begin{figure*}[t!]
\includegraphics{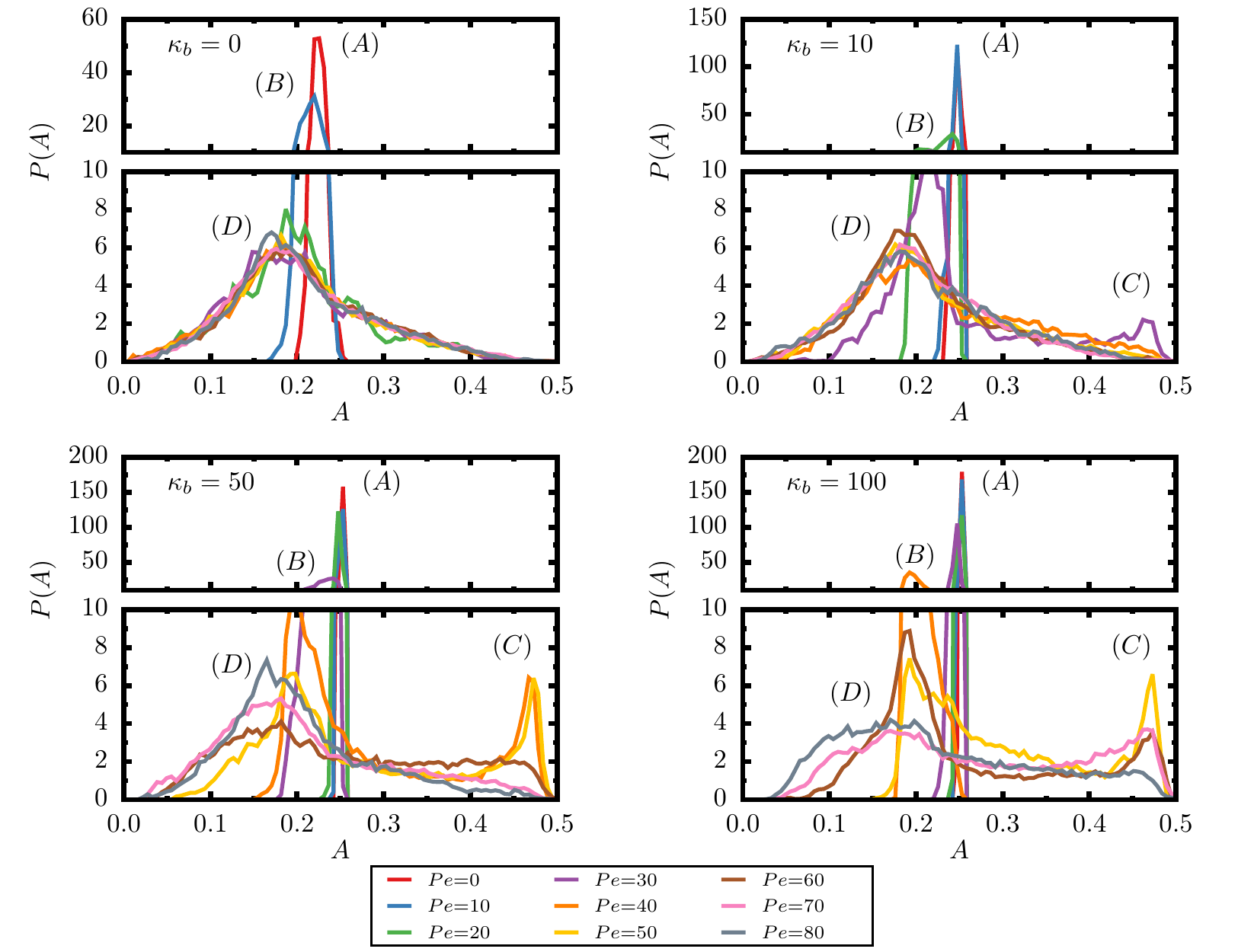}
\caption{\label{A} {(Color online) Probability distribution of the asphericity $P(A)$ for membranes of increasingly large Peclet numbers.  Different panels refer to different bending rigidities of the membrane as indicated.
The letters in parenthesis indicate the corresponding configurations as shown in Fig.~\ref{schematic}.
}}
\end{figure*}

We begin by considering the time average of the shape descriptors $\langle A \rangle$ and $\langle R_g \rangle$ 
for different values of the Peclet number.  In the limit of low bath activity ($Pe \rightarrow 0$), the membrane is found in a characteristically extended state with $\langle A \rangle \approx 0.22-0.25$ and the radius of gyration at a maximum  $ \langle R_g \rangle \approx 16-17\sigma$, with the larger values associated with the stiffer membranes. Notice that, unlike its one-dimensional counterpart (the polymer), an elastic membrane exhibits an overall extended state even in the limit for $\kappa_b=0$, which makes the dependence of $R_g$ with $Pe$ rather different than what was observed for fully flexible polymers confined to two dimensions\cite{kaiser044903,harder194901}. In fact, as shown in Fig.~\ref{avg_Rg}, as the bath activity is increased,  $R_g$ systematically decreases to smaller values until a plateau is reached. Stiffer membranes require larger $Pe$ before  $R_g$ begins to decrease and the value it decays to increases with the strength of the active force.  The behavior of the average asphericity as a function of $Pe$, suggests a more complex structural landscape. While the asphericity for the fully flexible membrane is relatively well behaved where it undergoes an initial decreases followed by a gradual increase for larger $Pe$, the typical conformations acquired by the rigid membranes are strongly dependent on the value of the bending rigidity and $Pe$, as illustrated by the sharp changes in $\langle A \rangle$ particularly for intermediate active forces.  

\begin{figure}[t!]
\includegraphics{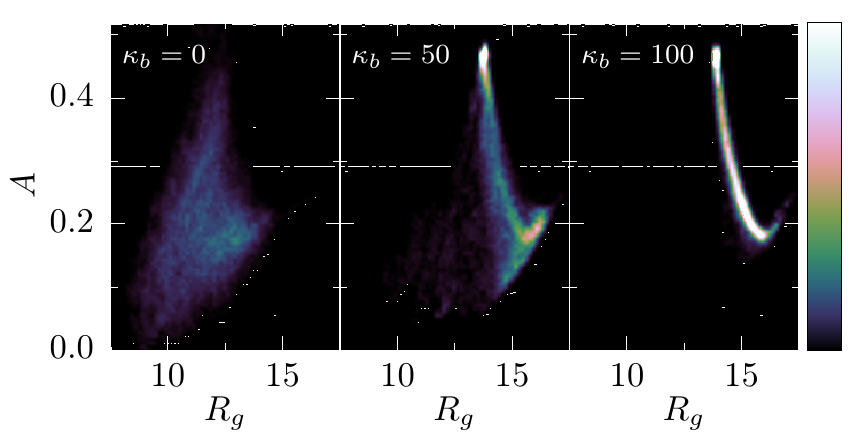}
\caption{\label{AvRg} {(Color online) Joint probability distribution of the membrane shape $P(A,R_g)$ for a fixed Peclet number $Pe=50$ and various bending rigidities.}}
\end{figure}

To gain additional insight into the shape of the membrane as a function of the active forces, we analyze the underlying probability distributions of the asphericity $P(A)$ and radius of gyration $P(R_g)$. The results are  given in Fig.~\ref{Rg} and Fig.~\ref{A}, respectively. 
For small active forces, as discussed above, the membrane is nearly flat and as expected the distributions for $P(A)$ and $P(R_g)$ are sharply peaked around the corresponding values of $R_g$ and $A$. In this low $Pe$ limit, the distributions become increasingly sharper as the bending rigidity is increased. For larger values of $Pe$,  both $P(R_g)$ and $P(A)$ broaden and shift towards smaller values for flexible membranes,
indicating that the membrane is on average more compact, but it can also access a variety of conformations across the spectrum
of possible shapes. More interestingly, for large bending rigidities distinct multiple peaks appear at specific values of $A$, which is a clear indications that the membrane breathes dynamically between a restricted number of partially stable conformations.  Specifically, at large bending rigidities and intermediate values of $Pe$, the membrane mainly interconverts between a bent configuration
(Fig.~\ref{schematic}B) characterized by $A\approx0.2$, and a configuration displaying a single fold along the center of the membrane
(Fig.~\ref{schematic}C) $A\approx0.48$. Visualization of the membrane trajectory (provided in the supporting materials) 
reveals a continuous folding and unfolding of the membrane over time.  For more flexible membranes, the dynamic behavior is similar, however, the most compact shape the membrane obtains is a multifolded configuration as shown in Fig.~\ref{schematic}D.  The larger degree of flexibility allows for a broader range of shape deformations under the forces generated by the active particles.  This is clearly observed in  Fig.~\ref{AvRg} where we plot the joint probability distribution $P (A, R_g )$ for a fixed Peclet number $Pe = 50$ and various bending rigidities. A flexible membranes explores a wide range of configurations with different $R_g$ and $A$, but as the stiffness of the membrane is increases the explorable landscape of shape conformations becomes more and more localized.

All of the conformations active particles induce on the membrane surface are the result of their tendency to localize near regions with large/positive curvature. The reasoning behind this phenomenon has been discussed in detail here \cite{Mallory032309,fily012125,fily5609,yang6477}.  A rigid membrane is nearly flat in the absence of active particles. When active particles are introduced, they tend to accumulate on the surface of the membrane. An initial bend of the membrane occurs once a sufficiently large asymmetry in the number of active particles develops on either side of its area.  This leads to an instability because the region of the membrane with higher/positive curvature will further stabilize the active particles residing on it, while the ones on the outer surface will be more easily escape. The net result is an increasing density gradient of particles on the surface.  Similarly to the mechanism driving the formation of hairpins in polymers \cite{harder062312,kaiser044903}, particles accumulating within the creased region of the membrane act as a dynamic fulcrum by which the sides of the membrane can pivot and fold onto one another.  Because no direct attractive interactions are present in our system, any folded or multi-folded configuration of the membrane will eventually unfold,
resulting in  repeating events of folding/unfolding transition, the rate of which increases with membrane flexibility.

Two  limits are worthy of a more detailed discussion.  The first case concerns the large Peclet number limit. In this limit, the probability distribution $P(A)$ associated to rigid membranes become comparable with that of their fully flexible counterparts. This is because larger active forces are able to fold the membrane simultaneously at different locations, leading to the formation of more compact multi-folds and to an overall softening of the bending rigidity.  Thus at large $Pe$ the dynamical behavior of a rigid membrane is nearly analogous to that of a flexible membrane.  The second limit concerns the case in which the membrane is sufficiently rigid that the active forces are not strong enough to drive the formation of a single fold. Here, the membrane  adopts partially bent configurations (Fig~\ref{schematic}B), and the surface acts as a sail able to trap a significant number of active particles preferentially on the side of the membrane with positive curvature. The net result is a  active motion of the surface
that is modulated by the propelling forces. The mechanism behind how an active fluid can induce an active transport on curved
micro-components as been discussed in our previous work~\cite{Mallory032309}.\\

\noindent \textit {Conclusions}-- 
In this paper we studied the behavior of a deformable elastic membrane in the presence of a low density  suspension 
of active particles as a function of the mechanical parameters of the membrane and the strength of the bath activity. 
We find that as soon as the collective strength of the  active forces becomes sufficiently large compared to the 
bending energy of the membrane a repeating sequence of folding and re-expansion transitions of the surface takes place. 
While flexible membranes tend to collapse into multi-folded states,  stiff membranes oscillates between an extended configuration and a singly folded state. Interestingly, in the large activity-limit, the behavior of rigid membranes resembles that of fully flexible ones, indicating 
that strong active forces can soften the modes of deformation of the membrane. Furthermore, we 
find that bent or partially folded membrane configurations act as sails or nets capable of trapping active particles on the positive curvatures of their 
surface and become actively transported through the medium.

This work, is complementary to our previous study on rigid filaments~\cite{Mallory032309}, and suggests 
ways of exploiting the random motion of active particles to perform mechanical work at the micro-scale. 
The system we considered here consisted of a simple  isotropic membrane in a bath of spherical particles; it
is feasible to imagine that further control over the mechanical response of flexible micro-components such as the ones 
discussed here and in our previous work can be achieved by introducing judicious amount of anisotropy either on the elastic properties of the surface or on the interactions between active particles and surface.\\

\noindent \textit {Acknowledgments}--We thank Clarion Tung and Joseph Harder for insightful discussions and helpful comments. AC acknowledges financial supported from the National Science Foundation under  Grant No. DMR-1408259. CV acknowledges financial support from a Ramon y Cajal tenure track, and from the National Project FIS2013-43209-P. SAM acknowledges financial support from the National Science Foundation Graduate Research Fellowship. This work used the Extreme Science and Engineering Discovery Environment (XSEDE), which is supported by National Science Foundation grant number ACI-1053575.

\bibliographystyle{apsrev4-1}

\bibliography{elastico}

\begin{thebibliography}{54}%
\makeatletter
\providecommand \@ifxundefined [1]{%
 \@ifx{#1\undefined}
}%
\providecommand \@ifnum [1]{%
 \ifnum #1\expandafter \@firstoftwo
 \else \expandafter \@secondoftwo
 \fi
}%
\providecommand \@ifx [1]{%
 \ifx #1\expandafter \@firstoftwo
 \else \expandafter \@secondoftwo
 \fi
}%
\providecommand \natexlab [1]{#1}%
\providecommand \enquote  [1]{``#1''}%
\providecommand \bibnamefont  [1]{#1}%
\providecommand \bibfnamefont [1]{#1}%
\providecommand \citenamefont [1]{#1}%
\providecommand \href@noop [0]{\@secondoftwo}%
\providecommand \href [0]{\begingroup \@sanitize@url \@href}%
\providecommand \@href[1]{\@@startlink{#1}\@@href}%
\providecommand \@@href[1]{\endgroup#1\@@endlink}%
\providecommand \@sanitize@url [0]{\catcode `\\12\catcode `\$12\catcode
  `\&12\catcode `\#12\catcode `\^12\catcode `\_12\catcode `\%12\relax}%
\providecommand \@@startlink[1]{}%
\providecommand \@@endlink[0]{}%
\providecommand \url  [0]{\begingroup\@sanitize@url \@url }%
\providecommand \@url [1]{\endgroup\@href {#1}{\urlprefix }}%
\providecommand \urlprefix  [0]{URL }%
\providecommand \Eprint [0]{\href }%
\providecommand \doibase [0]{http://dx.doi.org/}%
\providecommand \selectlanguage [0]{\@gobble}%
\providecommand \bibinfo  [0]{\@secondoftwo}%
\providecommand \bibfield  [0]{\@secondoftwo}%
\providecommand \translation [1]{[#1]}%
\providecommand \BibitemOpen [0]{}%
\providecommand \bibitemStop [0]{}%
\providecommand \bibitemNoStop [0]{.\EOS\space}%
\providecommand \EOS [0]{\spacefactor3000\relax}%
\providecommand \BibitemShut  [1]{\csname bibitem#1\endcsname}%
\let\auto@bib@innerbib\@empty
\bibitem [{\citenamefont {Wang}\ \emph {et~al.}(2013)\citenamefont {Wang},
  \citenamefont {Duan}, \citenamefont {Sen},\ and\ \citenamefont
  {Mallouk}}]{Wang17744}%
  \BibitemOpen
  \bibfield  {author} {\bibinfo {author} {\bibfnamefont {W.}~\bibnamefont
  {Wang}}, \bibinfo {author} {\bibfnamefont {W.}~\bibnamefont {Duan}}, \bibinfo
  {author} {\bibfnamefont {A.}~\bibnamefont {Sen}}, \ and\ \bibinfo {author}
  {\bibfnamefont {T.~E.}\ \bibnamefont {Mallouk}},\ }\href {\doibase
  10.1073/pnas.1311543110} {\bibfield  {journal} {\bibinfo  {journal} {Proc.
  Natl. Acad. Sci. USA}\ }\textbf {\bibinfo {volume} {110}},\ \bibinfo {pages}
  {17744} (\bibinfo {year} {2013})}\BibitemShut {NoStop}%
\bibitem [{\citenamefont {Leptos}\ \emph {et~al.}(2009)\citenamefont {Leptos},
  \citenamefont {Guasto}, \citenamefont {Gollub}, \citenamefont {Pesci},\ and\
  \citenamefont {Goldstein}}]{Leptos198103}%
  \BibitemOpen
  \bibfield  {author} {\bibinfo {author} {\bibfnamefont {K.~C.}\ \bibnamefont
  {Leptos}}, \bibinfo {author} {\bibfnamefont {J.~S.}\ \bibnamefont {Guasto}},
  \bibinfo {author} {\bibfnamefont {J.~P.}\ \bibnamefont {Gollub}}, \bibinfo
  {author} {\bibfnamefont {A.~I.}\ \bibnamefont {Pesci}}, \ and\ \bibinfo
  {author} {\bibfnamefont {R.~E.}\ \bibnamefont {Goldstein}},\ }\href {\doibase
  10.1103/PhysRevLett.103.198103} {\bibfield  {journal} {\bibinfo  {journal}
  {Phys. Rev. Lett.}\ }\textbf {\bibinfo {volume} {103}},\ \bibinfo {pages}
  {198103} (\bibinfo {year} {2009})}\BibitemShut {NoStop}%
\bibitem [{\citenamefont {Dunkel}\ \emph {et~al.}(2010)\citenamefont {Dunkel},
  \citenamefont {Putz}, \citenamefont {Zaid},\ and\ \citenamefont
  {Yeomans}}]{Dunkel4268}%
  \BibitemOpen
  \bibfield  {author} {\bibinfo {author} {\bibfnamefont {J.}~\bibnamefont
  {Dunkel}}, \bibinfo {author} {\bibfnamefont {V.~B.}\ \bibnamefont {Putz}},
  \bibinfo {author} {\bibfnamefont {I.~M.}\ \bibnamefont {Zaid}}, \ and\
  \bibinfo {author} {\bibfnamefont {J.~M.}\ \bibnamefont {Yeomans}},\ }\href
  {\doibase 10.1039/C0SM00164C} {\bibfield  {journal} {\bibinfo  {journal}
  {Soft Matter}\ }\textbf {\bibinfo {volume} {6}},\ \bibinfo {pages} {4268}
  (\bibinfo {year} {2010})}\BibitemShut {NoStop}%
\bibitem [{\citenamefont {Morozov}\ and\ \citenamefont
  {Marenduzzo}(2014)}]{Morozov2748}%
  \BibitemOpen
  \bibfield  {author} {\bibinfo {author} {\bibfnamefont {A.}~\bibnamefont
  {Morozov}}\ and\ \bibinfo {author} {\bibfnamefont {D.}~\bibnamefont
  {Marenduzzo}},\ }\href {\doibase 10.1039/C3SM52201F} {\bibfield  {journal}
  {\bibinfo  {journal} {Soft Matter}\ }\textbf {\bibinfo {volume} {10}},\
  \bibinfo {pages} {2748} (\bibinfo {year} {2014})}\BibitemShut {NoStop}%
\bibitem [{\citenamefont {Mallory}\ \emph
  {et~al.}(2014{\natexlab{a}})\citenamefont {Mallory}, \citenamefont
  {Valeriani},\ and\ \citenamefont {Cacciuto}}]{Mallory032309}%
  \BibitemOpen
  \bibfield  {author} {\bibinfo {author} {\bibfnamefont {S.~A.}\ \bibnamefont
  {Mallory}}, \bibinfo {author} {\bibfnamefont {C.}~\bibnamefont {Valeriani}},
  \ and\ \bibinfo {author} {\bibfnamefont {A.}~\bibnamefont {Cacciuto}},\
  }\href {\doibase 10.1103/PhysRevE.90.032309} {\bibfield  {journal} {\bibinfo
  {journal} {Phys. Rev. E}\ }\textbf {\bibinfo {volume} {90}},\ \bibinfo
  {pages} {032309} (\bibinfo {year} {2014}{\natexlab{a}})}\BibitemShut
  {NoStop}%
\bibitem [{\citenamefont {Molina}\ and\ \citenamefont
  {Yamamoto}(2014)}]{molina1389}%
  \BibitemOpen
  \bibfield  {author} {\bibinfo {author} {\bibfnamefont {J.~J.}\ \bibnamefont
  {Molina}}\ and\ \bibinfo {author} {\bibfnamefont {R.}~\bibnamefont
  {Yamamoto}},\ }\href@noop {} {\bibfield  {journal} {\bibinfo  {journal} {Mol.
  Phys.}\ }\textbf {\bibinfo {volume} {112}},\ \bibinfo {pages} {1389}
  (\bibinfo {year} {2014})}\BibitemShut {NoStop}%
\bibitem [{\citenamefont {Douglass}\ \emph {et~al.}(2012)\citenamefont
  {Douglass}, \citenamefont {Sukhov},\ and\ \citenamefont
  {Dogariu}}]{douglass834}%
  \BibitemOpen
  \bibfield  {author} {\bibinfo {author} {\bibfnamefont {K.~M.}\ \bibnamefont
  {Douglass}}, \bibinfo {author} {\bibfnamefont {S.}~\bibnamefont {Sukhov}}, \
  and\ \bibinfo {author} {\bibfnamefont {A.}~\bibnamefont {Dogariu}},\
  }\href@noop {} {\bibfield  {journal} {\bibinfo  {journal} {Nature Photon.}\
  }\textbf {\bibinfo {volume} {6}},\ \bibinfo {pages} {834} (\bibinfo {year}
  {2012})}\BibitemShut {NoStop}%
\bibitem [{\citenamefont {Miño}\ \emph {et~al.}(2013)\citenamefont {Miño},
  \citenamefont {Dunstan}, \citenamefont {Rousselet}, \citenamefont
  {Clément},\ and\ \citenamefont {Soto}}]{mino1469}%
  \BibitemOpen
  \bibfield  {author} {\bibinfo {author} {\bibfnamefont {G.~L.}\ \bibnamefont
  {Miño}}, \bibinfo {author} {\bibfnamefont {J.}~\bibnamefont {Dunstan}},
  \bibinfo {author} {\bibfnamefont {A.}~\bibnamefont {Rousselet}}, \bibinfo
  {author} {\bibfnamefont {E.}~\bibnamefont {Clément}}, \ and\ \bibinfo
  {author} {\bibfnamefont {R.}~\bibnamefont {Soto}},\ }\href {\doibase
  10.1017/jfm.2013.304} {\bibfield  {journal} {\bibinfo  {journal} {J. Fluid
  Mech.}\ }\textbf {\bibinfo {volume} {729}},\ \bibinfo {pages} {423} (\bibinfo
  {year} {2013})}\BibitemShut {NoStop}%
\bibitem [{\citenamefont {Kasyap}\ \emph {et~al.}(2014)\citenamefont {Kasyap},
  \citenamefont {Koch},\ and\ \citenamefont {Wu}}]{kasyap081901}%
  \BibitemOpen
  \bibfield  {author} {\bibinfo {author} {\bibfnamefont {T.~V.}\ \bibnamefont
  {Kasyap}}, \bibinfo {author} {\bibfnamefont {D.~L.}\ \bibnamefont {Koch}}, \
  and\ \bibinfo {author} {\bibfnamefont {M.}~\bibnamefont {Wu}},\ }\href
  {\doibase http://dx.doi.org/10.1063/1.4891570} {\bibfield  {journal}
  {\bibinfo  {journal} {Phys. Fluids}\ }\textbf {\bibinfo {volume} {26}},\
  \bibinfo {eid} {081901} (\bibinfo {year} {2014})}\BibitemShut {NoStop}%
\bibitem [{\citenamefont {Kaiser}\ \emph
  {et~al.}(2014{\natexlab{a}})\citenamefont {Kaiser}, \citenamefont {Peshkov},
  \citenamefont {Sokolov}, \citenamefont {ten Hagen}, \citenamefont {L\"owen},\
  and\ \citenamefont {Aranson}}]{kaiser158101}%
  \BibitemOpen
  \bibfield  {author} {\bibinfo {author} {\bibfnamefont {A.}~\bibnamefont
  {Kaiser}}, \bibinfo {author} {\bibfnamefont {A.}~\bibnamefont {Peshkov}},
  \bibinfo {author} {\bibfnamefont {A.}~\bibnamefont {Sokolov}}, \bibinfo
  {author} {\bibfnamefont {B.}~\bibnamefont {ten Hagen}}, \bibinfo {author}
  {\bibfnamefont {H.}~\bibnamefont {L\"owen}}, \ and\ \bibinfo {author}
  {\bibfnamefont {I.~S.}\ \bibnamefont {Aranson}},\ }\href {\doibase
  10.1103/PhysRevLett.112.158101} {\bibfield  {journal} {\bibinfo  {journal}
  {Phys. Rev. Lett.}\ }\textbf {\bibinfo {volume} {112}},\ \bibinfo {pages}
  {158101} (\bibinfo {year} {2014}{\natexlab{a}})}\BibitemShut {NoStop}%
\bibitem [{\citenamefont {Wu}\ and\ \citenamefont {Libchaber}(2000)}]{wu3017}%
  \BibitemOpen
  \bibfield  {author} {\bibinfo {author} {\bibfnamefont {X.-L.}\ \bibnamefont
  {Wu}}\ and\ \bibinfo {author} {\bibfnamefont {A.}~\bibnamefont {Libchaber}},\
  }\href {\doibase 10.1103/PhysRevLett.84.3017} {\bibfield  {journal} {\bibinfo
   {journal} {Phys. Rev. Lett.}\ }\textbf {\bibinfo {volume} {84}},\ \bibinfo
  {pages} {3017} (\bibinfo {year} {2000})}\BibitemShut {NoStop}%
\bibitem [{\citenamefont {Mi\~no}\ \emph {et~al.}(2011)\citenamefont {Mi\~no},
  \citenamefont {Mallouk}, \citenamefont {Darnige}, \citenamefont {Hoyos},
  \citenamefont {Dauchet}, \citenamefont {Dunstan}, \citenamefont {Soto},
  \citenamefont {Wang}, \citenamefont {Rousselet},\ and\ \citenamefont
  {Clement}}]{mino048102}%
  \BibitemOpen
  \bibfield  {author} {\bibinfo {author} {\bibfnamefont {G.}~\bibnamefont
  {Mi\~no}}, \bibinfo {author} {\bibfnamefont {T.~E.}\ \bibnamefont {Mallouk}},
  \bibinfo {author} {\bibfnamefont {T.}~\bibnamefont {Darnige}}, \bibinfo
  {author} {\bibfnamefont {M.}~\bibnamefont {Hoyos}}, \bibinfo {author}
  {\bibfnamefont {J.}~\bibnamefont {Dauchet}}, \bibinfo {author} {\bibfnamefont
  {J.}~\bibnamefont {Dunstan}}, \bibinfo {author} {\bibfnamefont
  {R.}~\bibnamefont {Soto}}, \bibinfo {author} {\bibfnamefont {Y.}~\bibnamefont
  {Wang}}, \bibinfo {author} {\bibfnamefont {A.}~\bibnamefont {Rousselet}}, \
  and\ \bibinfo {author} {\bibfnamefont {E.}~\bibnamefont {Clement}},\ }\href
  {\doibase 10.1103/PhysRevLett.106.048102} {\bibfield  {journal} {\bibinfo
  {journal} {Phys. Rev. Lett.}\ }\textbf {\bibinfo {volume} {106}},\ \bibinfo
  {pages} {048102} (\bibinfo {year} {2011})}\BibitemShut {NoStop}%
\bibitem [{\citenamefont {Eckhardt}\ and\ \citenamefont
  {Zammert}(2012)}]{eckhardt96}%
  \BibitemOpen
  \bibfield  {author} {\bibinfo {author} {\bibfnamefont {B.}~\bibnamefont
  {Eckhardt}}\ and\ \bibinfo {author} {\bibfnamefont {S.}~\bibnamefont
  {Zammert}},\ }\href {http://dx.doi.org/10.1140/epje/i2012-12096-7} {\bibfield
   {journal} {\bibinfo  {journal} {Eur. Phys. J. E Soft Matter}\ }\textbf
  {\bibinfo {volume} {35}},\ \bibinfo {eid} {96} (\bibinfo {year}
  {2012})}\BibitemShut {NoStop}%
\bibitem [{\citenamefont {Jepson}\ \emph {et~al.}(2013)\citenamefont {Jepson},
  \citenamefont {Martinez}, \citenamefont {Schwarz-Linek}, \citenamefont
  {Morozov},\ and\ \citenamefont {Poon}}]{jepson041002}%
  \BibitemOpen
  \bibfield  {author} {\bibinfo {author} {\bibfnamefont {A.}~\bibnamefont
  {Jepson}}, \bibinfo {author} {\bibfnamefont {V.~A.}\ \bibnamefont
  {Martinez}}, \bibinfo {author} {\bibfnamefont {J.}~\bibnamefont
  {Schwarz-Linek}}, \bibinfo {author} {\bibfnamefont {A.}~\bibnamefont
  {Morozov}}, \ and\ \bibinfo {author} {\bibfnamefont {W.~C.~K.}\ \bibnamefont
  {Poon}},\ }\href {\doibase 10.1103/PhysRevE.88.041002} {\bibfield  {journal}
  {\bibinfo  {journal} {Phys. Rev. E}\ }\textbf {\bibinfo {volume} {88}},\
  \bibinfo {pages} {041002} (\bibinfo {year} {2013})}\BibitemShut {NoStop}%
\bibitem [{\citenamefont {Shklarsh}\ \emph {et~al.}(2012)\citenamefont
  {Shklarsh}, \citenamefont {Finkelshtein}, \citenamefont {Ariel},
  \citenamefont {Kalisman}, \citenamefont {Ingham},\ and\ \citenamefont
  {Ben-Jacob}}]{shklarsh786}%
  \BibitemOpen
  \bibfield  {author} {\bibinfo {author} {\bibfnamefont {A.}~\bibnamefont
  {Shklarsh}}, \bibinfo {author} {\bibfnamefont {A.}~\bibnamefont
  {Finkelshtein}}, \bibinfo {author} {\bibfnamefont {G.}~\bibnamefont {Ariel}},
  \bibinfo {author} {\bibfnamefont {O.}~\bibnamefont {Kalisman}}, \bibinfo
  {author} {\bibfnamefont {C.}~\bibnamefont {Ingham}}, \ and\ \bibinfo {author}
  {\bibfnamefont {E.}~\bibnamefont {Ben-Jacob}},\ }\href {\doibase
  10.1098/rsfs.2012.0029} {\bibfield  {journal} {\bibinfo  {journal} {Interface
  Focus}\ }\textbf {\bibinfo {volume} {2}},\ \bibinfo {pages} {786} (\bibinfo
  {year} {2012})}\BibitemShut {NoStop}%
\bibitem [{\citenamefont {Angelani}\ and\ \citenamefont
  {Leonardo}(2010)}]{angelani113017}%
  \BibitemOpen
  \bibfield  {author} {\bibinfo {author} {\bibfnamefont {L.}~\bibnamefont
  {Angelani}}\ and\ \bibinfo {author} {\bibfnamefont {R.~D.}\ \bibnamefont
  {Leonardo}},\ }\href {http://stacks.iop.org/1367-2630/12/i=11/a=113017}
  {\bibfield  {journal} {\bibinfo  {journal} {New J. Phys.}\ }\textbf {\bibinfo
  {volume} {12}},\ \bibinfo {pages} {113017} (\bibinfo {year}
  {2010})}\BibitemShut {NoStop}%
\bibitem [{\citenamefont {Takagi}\ \emph {et~al.}(2014)\citenamefont {Takagi},
  \citenamefont {Palacci}, \citenamefont {Braunschweig}, \citenamefont
  {Shelley},\ and\ \citenamefont {Zhang}}]{takagi1784}%
  \BibitemOpen
  \bibfield  {author} {\bibinfo {author} {\bibfnamefont {D.}~\bibnamefont
  {Takagi}}, \bibinfo {author} {\bibfnamefont {J.}~\bibnamefont {Palacci}},
  \bibinfo {author} {\bibfnamefont {A.~B.}\ \bibnamefont {Braunschweig}},
  \bibinfo {author} {\bibfnamefont {M.~J.}\ \bibnamefont {Shelley}}, \ and\
  \bibinfo {author} {\bibfnamefont {J.}~\bibnamefont {Zhang}},\ }\href
  {\doibase 10.1039/C3SM52815D} {\bibfield  {journal} {\bibinfo  {journal}
  {Soft Matter}\ }\textbf {\bibinfo {volume} {10}},\ \bibinfo {pages} {1784}
  (\bibinfo {year} {2014})}\BibitemShut {NoStop}%
\bibitem [{\citenamefont {Kaiser}\ \emph
  {et~al.}(2014{\natexlab{b}})\citenamefont {Kaiser}, \citenamefont {Sokolov},
  \citenamefont {Aranson},\ and\ \citenamefont {L{\"o}wen}}]{kaiser1411.0964}%
  \BibitemOpen
  \bibfield  {author} {\bibinfo {author} {\bibfnamefont {A.}~\bibnamefont
  {Kaiser}}, \bibinfo {author} {\bibfnamefont {A.}~\bibnamefont {Sokolov}},
  \bibinfo {author} {\bibfnamefont {I.~S.}\ \bibnamefont {Aranson}}, \ and\
  \bibinfo {author} {\bibfnamefont {H.}~\bibnamefont {L{\"o}wen}},\ }\href@noop
  {} {\bibfield  {journal} {\bibinfo  {journal} {arXiv preprint
  arXiv:1411.0964}\ } (\bibinfo {year} {2014}{\natexlab{b}})}\BibitemShut
  {NoStop}%
\bibitem [{\citenamefont {Harder}\ \emph
  {et~al.}(2014{\natexlab{a}})\citenamefont {Harder}, \citenamefont {Mallory},
  \citenamefont {Tung}, \citenamefont {Valeriani},\ and\ \citenamefont
  {Cacciuto}}]{harder194901}%
  \BibitemOpen
  \bibfield  {author} {\bibinfo {author} {\bibfnamefont {J.}~\bibnamefont
  {Harder}}, \bibinfo {author} {\bibfnamefont {S.~A.}\ \bibnamefont {Mallory}},
  \bibinfo {author} {\bibfnamefont {C.}~\bibnamefont {Tung}}, \bibinfo {author}
  {\bibfnamefont {C.}~\bibnamefont {Valeriani}}, \ and\ \bibinfo {author}
  {\bibfnamefont {A.}~\bibnamefont {Cacciuto}},\ }\href {\doibase
  http://dx.doi.org/10.1063/1.4900720} {\bibfield  {journal} {\bibinfo
  {journal} {J. Chem. Phys.}\ }\textbf {\bibinfo {volume} {141}},\ \bibinfo
  {eid} {194901} (\bibinfo {year} {2014}{\natexlab{a}})}\BibitemShut {NoStop}%
\bibitem [{\citenamefont {Ni}\ \emph {et~al.}(2015)\citenamefont {Ni},
  \citenamefont {Cohen~Stuart},\ and\ \citenamefont {Bolhuis}}]{ni018302}%
  \BibitemOpen
  \bibfield  {author} {\bibinfo {author} {\bibfnamefont {R.}~\bibnamefont
  {Ni}}, \bibinfo {author} {\bibfnamefont {M.~A.}\ \bibnamefont
  {Cohen~Stuart}}, \ and\ \bibinfo {author} {\bibfnamefont {P.~G.}\
  \bibnamefont {Bolhuis}},\ }\href {\doibase 10.1103/PhysRevLett.114.018302}
  {\bibfield  {journal} {\bibinfo  {journal} {Phys. Rev. Lett.}\ }\textbf
  {\bibinfo {volume} {114}},\ \bibinfo {pages} {018302} (\bibinfo {year}
  {2015})}\BibitemShut {NoStop}%
\bibitem [{\citenamefont {Ray}\ \emph {et~al.}(2014)\citenamefont {Ray},
  \citenamefont {Reichhardt},\ and\ \citenamefont {Reichhardt}}]{ray013019}%
  \BibitemOpen
  \bibfield  {author} {\bibinfo {author} {\bibfnamefont {D.}~\bibnamefont
  {Ray}}, \bibinfo {author} {\bibfnamefont {C.}~\bibnamefont {Reichhardt}}, \
  and\ \bibinfo {author} {\bibfnamefont {C.~J.~O.}\ \bibnamefont
  {Reichhardt}},\ }\href {\doibase 10.1103/PhysRevE.90.013019} {\bibfield
  {journal} {\bibinfo  {journal} {Phys. Rev. E}\ }\textbf {\bibinfo {volume}
  {90}},\ \bibinfo {pages} {013019} (\bibinfo {year} {2014})}\BibitemShut
  {NoStop}%
\bibitem [{\citenamefont {Angelani}\ \emph {et~al.}(2011)\citenamefont
  {Angelani}, \citenamefont {Maggi}, \citenamefont {Bernardini}, \citenamefont
  {Rizzo},\ and\ \citenamefont {Di~Leonardo}}]{angelani138302}%
  \BibitemOpen
  \bibfield  {author} {\bibinfo {author} {\bibfnamefont {L.}~\bibnamefont
  {Angelani}}, \bibinfo {author} {\bibfnamefont {C.}~\bibnamefont {Maggi}},
  \bibinfo {author} {\bibfnamefont {M.~L.}\ \bibnamefont {Bernardini}},
  \bibinfo {author} {\bibfnamefont {A.}~\bibnamefont {Rizzo}}, \ and\ \bibinfo
  {author} {\bibfnamefont {R.}~\bibnamefont {Di~Leonardo}},\ }\href {\doibase
  10.1103/PhysRevLett.107.138302} {\bibfield  {journal} {\bibinfo  {journal}
  {Phys. Rev. Lett.}\ }\textbf {\bibinfo {volume} {107}},\ \bibinfo {pages}
  {138302} (\bibinfo {year} {2011})}\BibitemShut {NoStop}%
\bibitem [{\citenamefont {Das}\ \emph {et~al.}(2014)\citenamefont {Das},
  \citenamefont {Egorov}, \citenamefont {Trefz}, \citenamefont {Virnau},\ and\
  \citenamefont {Binder}}]{das198301}%
  \BibitemOpen
  \bibfield  {author} {\bibinfo {author} {\bibfnamefont {S.~K.}\ \bibnamefont
  {Das}}, \bibinfo {author} {\bibfnamefont {S.~A.}\ \bibnamefont {Egorov}},
  \bibinfo {author} {\bibfnamefont {B.}~\bibnamefont {Trefz}}, \bibinfo
  {author} {\bibfnamefont {P.}~\bibnamefont {Virnau}}, \ and\ \bibinfo {author}
  {\bibfnamefont {K.}~\bibnamefont {Binder}},\ }\href {\doibase
  10.1103/PhysRevLett.112.198301} {\bibfield  {journal} {\bibinfo  {journal}
  {Phys. Rev. Lett.}\ }\textbf {\bibinfo {volume} {112}},\ \bibinfo {pages}
  {198301} (\bibinfo {year} {2014})}\BibitemShut {NoStop}%
\bibitem [{\citenamefont {Krafnick}\ and\ \citenamefont
  {Garc\'ia}(2015)}]{krafnick022308}%
  \BibitemOpen
  \bibfield  {author} {\bibinfo {author} {\bibfnamefont {R.~C.}\ \bibnamefont
  {Krafnick}}\ and\ \bibinfo {author} {\bibfnamefont {A.~E.}\ \bibnamefont
  {Garc\'ia}},\ }\href {\doibase 10.1103/PhysRevE.91.022308} {\bibfield
  {journal} {\bibinfo  {journal} {Phys. Rev. E}\ }\textbf {\bibinfo {volume}
  {91}},\ \bibinfo {pages} {022308} (\bibinfo {year} {2015})}\BibitemShut
  {NoStop}%
\bibitem [{\citenamefont {Angelani}\ \emph {et~al.}(2009)\citenamefont
  {Angelani}, \citenamefont {Di~Leonardo},\ and\ \citenamefont
  {Ruocco}}]{Angelani048104}%
  \BibitemOpen
  \bibfield  {author} {\bibinfo {author} {\bibfnamefont {L.}~\bibnamefont
  {Angelani}}, \bibinfo {author} {\bibfnamefont {R.}~\bibnamefont
  {Di~Leonardo}}, \ and\ \bibinfo {author} {\bibfnamefont {G.}~\bibnamefont
  {Ruocco}},\ }\href {\doibase 10.1103/PhysRevLett.102.048104} {\bibfield
  {journal} {\bibinfo  {journal} {Phys. Rev. Lett.}\ }\textbf {\bibinfo
  {volume} {102}},\ \bibinfo {pages} {048104} (\bibinfo {year}
  {2009})}\BibitemShut {NoStop}%
\bibitem [{\citenamefont {Lambert}\ \emph {et~al.}(2010)\citenamefont
  {Lambert}, \citenamefont {Liao},\ and\ \citenamefont
  {Austin}}]{lambert168102}%
  \BibitemOpen
  \bibfield  {author} {\bibinfo {author} {\bibfnamefont {G.}~\bibnamefont
  {Lambert}}, \bibinfo {author} {\bibfnamefont {D.}~\bibnamefont {Liao}}, \
  and\ \bibinfo {author} {\bibfnamefont {R.~H.}\ \bibnamefont {Austin}},\
  }\href {\doibase 10.1103/PhysRevLett.104.168102} {\bibfield  {journal}
  {\bibinfo  {journal} {Phys. Rev. Lett.}\ }\textbf {\bibinfo {volume} {104}},\
  \bibinfo {pages} {168102} (\bibinfo {year} {2010})}\BibitemShut {NoStop}%
\bibitem [{\citenamefont {Sokolov}\ \emph {et~al.}(2010)\citenamefont
  {Sokolov}, \citenamefont {Apodaca}, \citenamefont {Grzybowski},\ and\
  \citenamefont {Aranson}}]{Sokolov969}%
  \BibitemOpen
  \bibfield  {author} {\bibinfo {author} {\bibfnamefont {A.}~\bibnamefont
  {Sokolov}}, \bibinfo {author} {\bibfnamefont {M.~M.}\ \bibnamefont
  {Apodaca}}, \bibinfo {author} {\bibfnamefont {B.~A.}\ \bibnamefont
  {Grzybowski}}, \ and\ \bibinfo {author} {\bibfnamefont {I.~S.}\ \bibnamefont
  {Aranson}},\ }\href {\doibase 10.1073/pnas.0913015107} {\bibfield  {journal}
  {\bibinfo  {journal} {Proc. Natl. Acad. Sci. USA}\ }\textbf {\bibinfo
  {volume} {107}},\ \bibinfo {pages} {969} (\bibinfo {year}
  {2010})}\BibitemShut {NoStop}%
\bibitem [{\citenamefont {Di~Leonardo}\ \emph {et~al.}(2010)\citenamefont
  {Di~Leonardo}, \citenamefont {Angelani}, \citenamefont {Dell’Arciprete},
  \citenamefont {Ruocco}, \citenamefont {Iebba}, \citenamefont {Schippa},
  \citenamefont {Conte}, \citenamefont {Mecarini}, \citenamefont {De~Angelis},\
  and\ \citenamefont {Di~Fabrizio}}]{DiLeonardo9541}%
  \BibitemOpen
  \bibfield  {author} {\bibinfo {author} {\bibfnamefont {R.}~\bibnamefont
  {Di~Leonardo}}, \bibinfo {author} {\bibfnamefont {L.}~\bibnamefont
  {Angelani}}, \bibinfo {author} {\bibfnamefont {D.}~\bibnamefont
  {Dell’Arciprete}}, \bibinfo {author} {\bibfnamefont {G.}~\bibnamefont
  {Ruocco}}, \bibinfo {author} {\bibfnamefont {V.}~\bibnamefont {Iebba}},
  \bibinfo {author} {\bibfnamefont {S.}~\bibnamefont {Schippa}}, \bibinfo
  {author} {\bibfnamefont {M.~P.}\ \bibnamefont {Conte}}, \bibinfo {author}
  {\bibfnamefont {F.}~\bibnamefont {Mecarini}}, \bibinfo {author}
  {\bibfnamefont {F.}~\bibnamefont {De~Angelis}}, \ and\ \bibinfo {author}
  {\bibfnamefont {E.}~\bibnamefont {Di~Fabrizio}},\ }\href {\doibase
  10.1073/pnas.0910426107} {\bibfield  {journal} {\bibinfo  {journal} {Proc.
  Natl. Acad. Sci. USA}\ }\textbf {\bibinfo {volume} {107}},\ \bibinfo {pages}
  {9541} (\bibinfo {year} {2010})}\BibitemShut {NoStop}%
\bibitem [{\citenamefont {Koumakis}\ \emph {et~al.}(2013)\citenamefont
  {Koumakis}, \citenamefont {Lepore}, \citenamefont {Maggi},\ and\
  \citenamefont {Di~Leonardo}}]{koumakis}%
  \BibitemOpen
  \bibfield  {author} {\bibinfo {author} {\bibfnamefont {N.}~\bibnamefont
  {Koumakis}}, \bibinfo {author} {\bibfnamefont {A.}~\bibnamefont {Lepore}},
  \bibinfo {author} {\bibfnamefont {C.}~\bibnamefont {Maggi}}, \ and\ \bibinfo
  {author} {\bibfnamefont {R.}~\bibnamefont {Di~Leonardo}},\ }\href@noop {}
  {\bibfield  {journal} {\bibinfo  {journal} {Nat. Commun.}\ }\textbf {\bibinfo
  {volume} {4}} (\bibinfo {year} {2013})}\BibitemShut {NoStop}%
\bibitem [{\citenamefont {Schwarz-Linek}\ \emph {et~al.}(2012)\citenamefont
  {Schwarz-Linek}, \citenamefont {Valeriani}, \citenamefont {Cacciuto},
  \citenamefont {Cates}, \citenamefont {Marenduzzo}, \citenamefont {Morozov},\
  and\ \citenamefont {Poon}}]{schwarz4052}%
  \BibitemOpen
  \bibfield  {author} {\bibinfo {author} {\bibfnamefont {J.}~\bibnamefont
  {Schwarz-Linek}}, \bibinfo {author} {\bibfnamefont {C.}~\bibnamefont
  {Valeriani}}, \bibinfo {author} {\bibfnamefont {A.}~\bibnamefont {Cacciuto}},
  \bibinfo {author} {\bibfnamefont {M.~E.}\ \bibnamefont {Cates}}, \bibinfo
  {author} {\bibfnamefont {D.}~\bibnamefont {Marenduzzo}}, \bibinfo {author}
  {\bibfnamefont {A.~N.}\ \bibnamefont {Morozov}}, \ and\ \bibinfo {author}
  {\bibfnamefont {W.~C.~K.}\ \bibnamefont {Poon}},\ }\href {\doibase
  10.1073/pnas.1116334109} {\bibfield  {journal} {\bibinfo  {journal} {Proc.
  Natl. Acad. Sci. USA}\ }\textbf {\bibinfo {volume} {109}},\ \bibinfo {pages}
  {4052} (\bibinfo {year} {2012})}\BibitemShut {NoStop}%
\bibitem [{\citenamefont {Grosberg}\ and\ \citenamefont
  {Joanny}(2015)}]{grosberg1502.08034}%
  \BibitemOpen
  \bibfield  {author} {\bibinfo {author} {\bibfnamefont {A.~Y.}\ \bibnamefont
  {Grosberg}}\ and\ \bibinfo {author} {\bibfnamefont {J.-F.}\ \bibnamefont
  {Joanny}},\ }\href@noop {} {\bibfield  {journal} {\bibinfo  {journal} {arXiv
  preprint arXiv:1502.08034}\ } (\bibinfo {year} {2015})}\BibitemShut {NoStop}%
\bibitem [{\citenamefont {Solon}\ \emph
  {et~al.}(2014{\natexlab{a}})\citenamefont {Solon}, \citenamefont {Fily},
  \citenamefont {Baskaran}, \citenamefont {Cates}, \citenamefont {Kafri},
  \citenamefont {Kardar},\ and\ \citenamefont {Tailleur}}]{solon1412.3952}%
  \BibitemOpen
  \bibfield  {author} {\bibinfo {author} {\bibfnamefont {A.}~\bibnamefont
  {Solon}}, \bibinfo {author} {\bibfnamefont {Y.}~\bibnamefont {Fily}},
  \bibinfo {author} {\bibfnamefont {A.}~\bibnamefont {Baskaran}}, \bibinfo
  {author} {\bibfnamefont {M.}~\bibnamefont {Cates}}, \bibinfo {author}
  {\bibfnamefont {Y.}~\bibnamefont {Kafri}}, \bibinfo {author} {\bibfnamefont
  {M.}~\bibnamefont {Kardar}}, \ and\ \bibinfo {author} {\bibfnamefont
  {J.}~\bibnamefont {Tailleur}},\ }\href@noop {} {\bibfield  {journal}
  {\bibinfo  {journal} {arXiv preprint arXiv:1412.3952}\ } (\bibinfo {year}
  {2014}{\natexlab{a}})}\BibitemShut {NoStop}%
\bibitem [{\citenamefont {Solon}\ \emph
  {et~al.}(2014{\natexlab{b}})\citenamefont {Solon}, \citenamefont
  {Stenhammar}, \citenamefont {Wittkowski}, \citenamefont {Kardar},
  \citenamefont {Kafri}, \citenamefont {Cates},\ and\ \citenamefont
  {Tailleur}}]{solon1412.5475}%
  \BibitemOpen
  \bibfield  {author} {\bibinfo {author} {\bibfnamefont {A.~P.}\ \bibnamefont
  {Solon}}, \bibinfo {author} {\bibfnamefont {J.}~\bibnamefont {Stenhammar}},
  \bibinfo {author} {\bibfnamefont {R.}~\bibnamefont {Wittkowski}}, \bibinfo
  {author} {\bibfnamefont {M.}~\bibnamefont {Kardar}}, \bibinfo {author}
  {\bibfnamefont {Y.}~\bibnamefont {Kafri}}, \bibinfo {author} {\bibfnamefont
  {M.~E.}\ \bibnamefont {Cates}}, \ and\ \bibinfo {author} {\bibfnamefont
  {J.}~\bibnamefont {Tailleur}},\ }\href@noop {} {\bibfield  {journal}
  {\bibinfo  {journal} {arXiv preprint arXiv:1412.5475}\ } (\bibinfo {year}
  {2014}{\natexlab{b}})}\BibitemShut {NoStop}%
\bibitem [{\citenamefont {Mallory}\ \emph
  {et~al.}(2014{\natexlab{b}})\citenamefont {Mallory}, \citenamefont {\ifmmode
  \check{S}\else \v{S}\fi{}ari\ifmmode~\acute{c}\else \'{c}\fi{}},
  \citenamefont {Valeriani},\ and\ \citenamefont {Cacciuto}}]{mallory052303}%
  \BibitemOpen
  \bibfield  {author} {\bibinfo {author} {\bibfnamefont {S.~A.}\ \bibnamefont
  {Mallory}}, \bibinfo {author} {\bibfnamefont {A.}~\bibnamefont {\ifmmode
  \check{S}\else \v{S}\fi{}ari\ifmmode~\acute{c}\else \'{c}\fi{}}}, \bibinfo
  {author} {\bibfnamefont {C.}~\bibnamefont {Valeriani}}, \ and\ \bibinfo
  {author} {\bibfnamefont {A.}~\bibnamefont {Cacciuto}},\ }\href {\doibase
  10.1103/PhysRevE.89.052303} {\bibfield  {journal} {\bibinfo  {journal} {Phys.
  Rev. E}\ }\textbf {\bibinfo {volume} {89}},\ \bibinfo {pages} {052303}
  (\bibinfo {year} {2014}{\natexlab{b}})}\BibitemShut {NoStop}%
\bibitem [{\citenamefont {Fily}\ \emph {et~al.}(2014)\citenamefont {Fily},
  \citenamefont {Baskaran},\ and\ \citenamefont {Hagan}}]{fily5609}%
  \BibitemOpen
  \bibfield  {author} {\bibinfo {author} {\bibfnamefont {Y.}~\bibnamefont
  {Fily}}, \bibinfo {author} {\bibfnamefont {A.}~\bibnamefont {Baskaran}}, \
  and\ \bibinfo {author} {\bibfnamefont {M.~F.}\ \bibnamefont {Hagan}},\ }\href
  {\doibase 10.1039/C4SM00975D} {\bibfield  {journal} {\bibinfo  {journal}
  {Soft Matter}\ }\textbf {\bibinfo {volume} {10}},\ \bibinfo {pages} {5609}
  (\bibinfo {year} {2014})}\BibitemShut {NoStop}%
\bibitem [{\citenamefont {Yang}\ \emph {et~al.}(2014)\citenamefont {Yang},
  \citenamefont {Manning},\ and\ \citenamefont {Marchetti}}]{yang6477}%
  \BibitemOpen
  \bibfield  {author} {\bibinfo {author} {\bibfnamefont {X.}~\bibnamefont
  {Yang}}, \bibinfo {author} {\bibfnamefont {M.~L.}\ \bibnamefont {Manning}}, \
  and\ \bibinfo {author} {\bibfnamefont {M.~C.}\ \bibnamefont {Marchetti}},\
  }\href {\doibase 10.1039/C4SM00927D} {\bibfield  {journal} {\bibinfo
  {journal} {Soft Matter}\ }\textbf {\bibinfo {volume} {10}},\ \bibinfo {pages}
  {6477} (\bibinfo {year} {2014})}\BibitemShut {NoStop}%
\bibitem [{\citenamefont {Takatori}\ and\ \citenamefont
  {Brady}(2014{\natexlab{a}})}]{takatori1411.5776}%
  \BibitemOpen
  \bibfield  {author} {\bibinfo {author} {\bibfnamefont {S.~C.}\ \bibnamefont
  {Takatori}}\ and\ \bibinfo {author} {\bibfnamefont {J.~F.}\ \bibnamefont
  {Brady}},\ }\href@noop {} {\bibfield  {journal} {\bibinfo  {journal} {arXiv
  preprint arXiv:1411.5776}\ } (\bibinfo {year}
  {2014}{\natexlab{a}})}\BibitemShut {NoStop}%
\bibitem [{\citenamefont {Ginot}\ \emph {et~al.}(2015)\citenamefont {Ginot},
  \citenamefont {Theurkauff}, \citenamefont {Levis}, \citenamefont {Ybert},
  \citenamefont {Bocquet}, \citenamefont {Berthier},\ and\ \citenamefont
  {Cottin-Bizonne}}]{ginot011004}%
  \BibitemOpen
  \bibfield  {author} {\bibinfo {author} {\bibfnamefont {F.}~\bibnamefont
  {Ginot}}, \bibinfo {author} {\bibfnamefont {I.}~\bibnamefont {Theurkauff}},
  \bibinfo {author} {\bibfnamefont {D.}~\bibnamefont {Levis}}, \bibinfo
  {author} {\bibfnamefont {C.}~\bibnamefont {Ybert}}, \bibinfo {author}
  {\bibfnamefont {L.}~\bibnamefont {Bocquet}}, \bibinfo {author} {\bibfnamefont
  {L.}~\bibnamefont {Berthier}}, \ and\ \bibinfo {author} {\bibfnamefont
  {C.}~\bibnamefont {Cottin-Bizonne}},\ }\href {\doibase
  10.1103/PhysRevX.5.011004} {\bibfield  {journal} {\bibinfo  {journal} {Phys.
  Rev. X}\ }\textbf {\bibinfo {volume} {5}},\ \bibinfo {pages} {011004}
  (\bibinfo {year} {2015})}\BibitemShut {NoStop}%
\bibitem [{\citenamefont {Takatori}\ and\ \citenamefont
  {Brady}(2014{\natexlab{b}})}]{takatori028103}%
  \BibitemOpen
  \bibfield  {author} {\bibinfo {author} {\bibfnamefont {S.~C.}\ \bibnamefont
  {Takatori}}\ and\ \bibinfo {author} {\bibfnamefont {J.~F.}\ \bibnamefont
  {Brady}},\ }\href {\doibase 10.1103/PhysRevLett.113.028103} {\bibfield
  {journal} {\bibinfo  {journal} {Phys. Rev. Lett.}\ }\textbf {\bibinfo
  {volume} {113}},\ \bibinfo {pages} {028103} (\bibinfo {year}
  {2014}{\natexlab{b}})}\BibitemShut {NoStop}%
\bibitem [{\citenamefont {Takatori}\ and\ \citenamefont
  {Brady}(2014{\natexlab{c}})}]{takatori9433}%
  \BibitemOpen
  \bibfield  {author} {\bibinfo {author} {\bibfnamefont {S.~C.}\ \bibnamefont
  {Takatori}}\ and\ \bibinfo {author} {\bibfnamefont {J.~F.}\ \bibnamefont
  {Brady}},\ }\href {\doibase 10.1039/C4SM01409J} {\bibfield  {journal}
  {\bibinfo  {journal} {Soft Matter}\ }\textbf {\bibinfo {volume} {10}},\
  \bibinfo {pages} {9433} (\bibinfo {year} {2014}{\natexlab{c}})}\BibitemShut
  {NoStop}%
\bibitem [{\citenamefont {Harder}\ \emph
  {et~al.}(2014{\natexlab{b}})\citenamefont {Harder}, \citenamefont
  {Valeriani},\ and\ \citenamefont {Cacciuto}}]{harder062312}%
  \BibitemOpen
  \bibfield  {author} {\bibinfo {author} {\bibfnamefont {J.}~\bibnamefont
  {Harder}}, \bibinfo {author} {\bibfnamefont {C.}~\bibnamefont {Valeriani}}, \
  and\ \bibinfo {author} {\bibfnamefont {A.}~\bibnamefont {Cacciuto}},\ }\href
  {\doibase 10.1103/PhysRevE.90.062312} {\bibfield  {journal} {\bibinfo
  {journal} {Phys. Rev. E}\ }\textbf {\bibinfo {volume} {90}},\ \bibinfo
  {pages} {062312} (\bibinfo {year} {2014}{\natexlab{b}})}\BibitemShut
  {NoStop}%
\bibitem [{\citenamefont {Kaiser}\ and\ \citenamefont
  {L{\"o}wen}(2014)}]{kaiser044903}%
  \BibitemOpen
  \bibfield  {author} {\bibinfo {author} {\bibfnamefont {A.}~\bibnamefont
  {Kaiser}}\ and\ \bibinfo {author} {\bibfnamefont {H.}~\bibnamefont
  {L{\"o}wen}},\ }\href {\doibase http://dx.doi.org/10.1063/1.4891095}
  {\bibfield  {journal} {\bibinfo  {journal} {J. Chem. Phys.}\ }\textbf
  {\bibinfo {volume} {141}},\ \bibinfo {eid} {044903} (\bibinfo {year}
  {2014})}\BibitemShut {NoStop}%
\bibitem [{\citenamefont {Meyer}\ \emph {et~al.}(2007)\citenamefont {Meyer},
  \citenamefont {Geim}, \citenamefont {Katsnelson}, \citenamefont {Novoselov},
  \citenamefont {Booth},\ and\ \citenamefont {Roth}}]{meyer60}%
  \BibitemOpen
  \bibfield  {author} {\bibinfo {author} {\bibfnamefont {J.~C.}\ \bibnamefont
  {Meyer}}, \bibinfo {author} {\bibfnamefont {A.~K.}\ \bibnamefont {Geim}},
  \bibinfo {author} {\bibfnamefont {M.}~\bibnamefont {Katsnelson}}, \bibinfo
  {author} {\bibfnamefont {K.}~\bibnamefont {Novoselov}}, \bibinfo {author}
  {\bibfnamefont {T.}~\bibnamefont {Booth}}, \ and\ \bibinfo {author}
  {\bibfnamefont {S.}~\bibnamefont {Roth}},\ }\href@noop {} {\bibfield
  {journal} {\bibinfo  {journal} {Nature}\ }\textbf {\bibinfo {volume} {446}},\
  \bibinfo {pages} {60} (\bibinfo {year} {2007})}\BibitemShut {NoStop}%
\bibitem [{\citenamefont {Stankovich}\ \emph {et~al.}(2006)\citenamefont
  {Stankovich}, \citenamefont {Dikin}, \citenamefont {Dommett}, \citenamefont
  {Kohlhaas}, \citenamefont {Zimney}, \citenamefont {Stach}, \citenamefont
  {Piner}, \citenamefont {Nguyen},\ and\ \citenamefont
  {Ruoff}}]{stankovich282}%
  \BibitemOpen
  \bibfield  {author} {\bibinfo {author} {\bibfnamefont {S.}~\bibnamefont
  {Stankovich}}, \bibinfo {author} {\bibfnamefont {D.~A.}\ \bibnamefont
  {Dikin}}, \bibinfo {author} {\bibfnamefont {G.~H.}\ \bibnamefont {Dommett}},
  \bibinfo {author} {\bibfnamefont {K.~M.}\ \bibnamefont {Kohlhaas}}, \bibinfo
  {author} {\bibfnamefont {E.~J.}\ \bibnamefont {Zimney}}, \bibinfo {author}
  {\bibfnamefont {E.~A.}\ \bibnamefont {Stach}}, \bibinfo {author}
  {\bibfnamefont {R.~D.}\ \bibnamefont {Piner}}, \bibinfo {author}
  {\bibfnamefont {S.~T.}\ \bibnamefont {Nguyen}}, \ and\ \bibinfo {author}
  {\bibfnamefont {R.~S.}\ \bibnamefont {Ruoff}},\ }\href@noop {} {\bibfield
  {journal} {\bibinfo  {journal} {Nature}\ }\textbf {\bibinfo {volume} {442}},\
  \bibinfo {pages} {282} (\bibinfo {year} {2006})}\BibitemShut {NoStop}%
\bibitem [{\citenamefont {Wen}\ \emph {et~al.}(1992)\citenamefont {Wen},
  \citenamefont {Garland}, \citenamefont {Hwa}, \citenamefont {Kardar},
  \citenamefont {Kokufuta}, \citenamefont {Li}, \citenamefont {Orkisz},\ and\
  \citenamefont {Tanaka}}]{wen426}%
  \BibitemOpen
  \bibfield  {author} {\bibinfo {author} {\bibfnamefont {X.}~\bibnamefont
  {Wen}}, \bibinfo {author} {\bibfnamefont {C.~W.}\ \bibnamefont {Garland}},
  \bibinfo {author} {\bibfnamefont {T.}~\bibnamefont {Hwa}}, \bibinfo {author}
  {\bibfnamefont {M.}~\bibnamefont {Kardar}}, \bibinfo {author} {\bibfnamefont
  {E.}~\bibnamefont {Kokufuta}}, \bibinfo {author} {\bibfnamefont
  {Y.}~\bibnamefont {Li}}, \bibinfo {author} {\bibfnamefont {M.}~\bibnamefont
  {Orkisz}}, \ and\ \bibinfo {author} {\bibfnamefont {T.}~\bibnamefont
  {Tanaka}},\ }\href@noop {} {\bibfield  {journal} {\bibinfo  {journal}
  {Nature}\ }\textbf {\bibinfo {volume} {355}},\ \bibinfo {pages} {426}
  (\bibinfo {year} {1992})}\BibitemShut {NoStop}%
\bibitem [{\citenamefont {Spector}\ \emph {et~al.}(1994)\citenamefont
  {Spector}, \citenamefont {Naranjo}, \citenamefont {Chiruvolu},\ and\
  \citenamefont {Zasadzinski}}]{spector2867}%
  \BibitemOpen
  \bibfield  {author} {\bibinfo {author} {\bibfnamefont {M.~S.}\ \bibnamefont
  {Spector}}, \bibinfo {author} {\bibfnamefont {E.}~\bibnamefont {Naranjo}},
  \bibinfo {author} {\bibfnamefont {S.}~\bibnamefont {Chiruvolu}}, \ and\
  \bibinfo {author} {\bibfnamefont {J.~A.}\ \bibnamefont {Zasadzinski}},\
  }\href {\doibase 10.1103/PhysRevLett.73.2867} {\bibfield  {journal} {\bibinfo
   {journal} {Phys. Rev. Lett.}\ }\textbf {\bibinfo {volume} {73}},\ \bibinfo
  {pages} {2867} (\bibinfo {year} {1994})}\BibitemShut {NoStop}%
\bibitem [{\citenamefont {Fendler}\ and\ \citenamefont
  {Tundo}(1984)}]{fendler3}%
  \BibitemOpen
  \bibfield  {author} {\bibinfo {author} {\bibfnamefont {J.~H.}\ \bibnamefont
  {Fendler}}\ and\ \bibinfo {author} {\bibfnamefont {P.}~\bibnamefont
  {Tundo}},\ }\href@noop {} {\bibfield  {journal} {\bibinfo  {journal} {Acc.
  Chem. Res.}\ }\textbf {\bibinfo {volume} {17}},\ \bibinfo {pages} {3}
  (\bibinfo {year} {1984})}\BibitemShut {NoStop}%
\bibitem [{\citenamefont {Elgsaeter}\ \emph {et~al.}(1986)\citenamefont
  {Elgsaeter}, \citenamefont {Stokke}, \citenamefont {Mikkelsen},\ and\
  \citenamefont {Branton}}]{elgsaeter1217}%
  \BibitemOpen
  \bibfield  {author} {\bibinfo {author} {\bibfnamefont {A.}~\bibnamefont
  {Elgsaeter}}, \bibinfo {author} {\bibfnamefont {B.~T.}\ \bibnamefont
  {Stokke}}, \bibinfo {author} {\bibfnamefont {A.}~\bibnamefont {Mikkelsen}}, \
  and\ \bibinfo {author} {\bibfnamefont {D.}~\bibnamefont {Branton}},\
  }\href@noop {} {\bibfield  {journal} {\bibinfo  {journal} {Science}\ }\textbf
  {\bibinfo {volume} {234}},\ \bibinfo {pages} {1217} (\bibinfo {year}
  {1986})}\BibitemShut {NoStop}%
\bibitem [{\citenamefont {Schmidt}\ \emph {et~al.}(1993)\citenamefont
  {Schmidt}, \citenamefont {Svoboda}, \citenamefont {Lei}, \citenamefont
  {Petsche}, \citenamefont {Berman}, \citenamefont {Safinya},\ and\
  \citenamefont {Grest}}]{schmidt952}%
  \BibitemOpen
  \bibfield  {author} {\bibinfo {author} {\bibfnamefont {C.~F.}\ \bibnamefont
  {Schmidt}}, \bibinfo {author} {\bibfnamefont {K.}~\bibnamefont {Svoboda}},
  \bibinfo {author} {\bibfnamefont {N.}~\bibnamefont {Lei}}, \bibinfo {author}
  {\bibfnamefont {I.~B.}\ \bibnamefont {Petsche}}, \bibinfo {author}
  {\bibfnamefont {L.~E.}\ \bibnamefont {Berman}}, \bibinfo {author}
  {\bibfnamefont {C.~R.}\ \bibnamefont {Safinya}}, \ and\ \bibinfo {author}
  {\bibfnamefont {G.~S.}\ \bibnamefont {Grest}},\ }\href@noop {} {\bibfield
  {journal} {\bibinfo  {journal} {Science}\ }\textbf {\bibinfo {volume}
  {259}},\ \bibinfo {pages} {952} (\bibinfo {year} {1993})}\BibitemShut
  {NoStop}%
\bibitem [{\citenamefont {Bowick}\ and\ \citenamefont
  {Travesset}(2001)}]{bowick255}%
  \BibitemOpen
  \bibfield  {author} {\bibinfo {author} {\bibfnamefont {M.~J.}\ \bibnamefont
  {Bowick}}\ and\ \bibinfo {author} {\bibfnamefont {A.}~\bibnamefont
  {Travesset}},\ }\href@noop {} {\bibfield  {journal} {\bibinfo  {journal}
  {Phys. Rep.}\ }\textbf {\bibinfo {volume} {344}},\ \bibinfo {pages} {255}
  (\bibinfo {year} {2001})}\BibitemShut {NoStop}%
\bibitem [{\citenamefont {Kantor}\ and\ \citenamefont
  {Nelson}(1987)}]{kantor2774}%
  \BibitemOpen
  \bibfield  {author} {\bibinfo {author} {\bibfnamefont {Y.}~\bibnamefont
  {Kantor}}\ and\ \bibinfo {author} {\bibfnamefont {D.~R.}\ \bibnamefont
  {Nelson}},\ }\href {\doibase 10.1103/PhysRevLett.58.2774} {\bibfield
  {journal} {\bibinfo  {journal} {Phys. Rev. Lett.}\ }\textbf {\bibinfo
  {volume} {58}},\ \bibinfo {pages} {2774} (\bibinfo {year}
  {1987})}\BibitemShut {NoStop}%
\bibitem [{\citenamefont {Plimpton}(1995)}]{plimpton1}%
  \BibitemOpen
  \bibfield  {author} {\bibinfo {author} {\bibfnamefont {S.}~\bibnamefont
  {Plimpton}},\ }\href@noop {} {\bibfield  {journal} {\bibinfo  {journal} {J.
  Comput. Phys.}\ }\textbf {\bibinfo {volume} {117}},\ \bibinfo {pages} {1}
  (\bibinfo {year} {1995})}\BibitemShut {NoStop}%
\bibitem [{\citenamefont {Rudnick}\ and\ \citenamefont
  {Gaspari}(1986)}]{rudnickL191}%
  \BibitemOpen
  \bibfield  {author} {\bibinfo {author} {\bibfnamefont {J.}~\bibnamefont
  {Rudnick}}\ and\ \bibinfo {author} {\bibfnamefont {G.}~\bibnamefont
  {Gaspari}},\ }\href {http://stacks.iop.org/0305-4470/19/i=4/a=004} {\bibfield
   {journal} {\bibinfo  {journal} {J. Phys. A}\ }\textbf {\bibinfo {volume}
  {19}},\ \bibinfo {pages} {L191} (\bibinfo {year} {1986})}\BibitemShut
  {NoStop}%
\bibitem [{\citenamefont {Fily}\ \emph {et~al.}(2015)\citenamefont {Fily},
  \citenamefont {Baskaran},\ and\ \citenamefont {Hagan}}]{fily012125}%
  \BibitemOpen
  \bibfield  {author} {\bibinfo {author} {\bibfnamefont {Y.}~\bibnamefont
  {Fily}}, \bibinfo {author} {\bibfnamefont {A.}~\bibnamefont {Baskaran}}, \
  and\ \bibinfo {author} {\bibfnamefont {M.~F.}\ \bibnamefont {Hagan}},\ }\href
  {\doibase 10.1103/PhysRevE.91.012125} {\bibfield  {journal} {\bibinfo
  {journal} {Phys. Rev. E}\ }\textbf {\bibinfo {volume} {91}},\ \bibinfo
  {pages} {012125} (\bibinfo {year} {2015})}\BibitemShut {NoStop}%
\end{thebibliography}%

\end{document}